\documentclass[fleqn,usenatbib]{mnras}

\usepackage[T1]{fontenc}
\usepackage{ae,aecompl}

\usepackage{graphicx}	
\usepackage{amsmath}	
\usepackage{amssymb}	
\usepackage{threeparttable}
\usepackage{tablefootnote}
\usepackage{newtxtext}
\usepackage[slantedGreek]{newtxmath}

\newcommand{\grb}{GRB 171010A}

\title[A radio study of GRB 171010A]{A detailed radio study of the energetic, nearby and puzzling GRB 171010A}

\author[J. S. Bright]{J. S. Bright,$^{1}$\thanks{E-mail: joe.bright@physics.ox.ac.uk} 
A. Horesh,$^{2}$ A. J. van der Horst$^{3,4}$, R. Fender,$^{1}$ G. E. Anderson,$^{5}$ \newauthor
S. E. Motta,$^{1}$ S. B. Cenko,$^{6}$ D. A. Green,$^{7}$ Y. Perrott,$^{7,8}$, D. Titterington$^{7}$
\\
$^{1}$Astrophysics, Department of Physics, Denys Wilkinson Building, Keble Road, Oxford OX1 3RH, UK\\
$^{2}$Racah Institute of Physics, The Hebrew University, Jerusalem 91904, Israel\\
$^{3}$Department of Physics, the George Washington University, 725 21\textsuperscript{st} Street NW, Washington, DC 20052, USA\\
$^{4}$Astronomy, Physics and Statistics Institute of Sciences (APSIS), 725 21\textsuperscript{st} Street NW, Washington, DC 20052, USA\\
$^{5}$International Centre for Radio Astronomy Research, Curtin University, GPO Box U1987, Perth, WA 6845, Australia\\
$^{6}$Astrophysics Science Division, NASA Goddard Space Flight Center, Mail Code 661, Greenbelt, MA 20771, USA\\
$^{7}$Astrophysics Group, Cavendish Laboratory, 19 J. J. Thomson Avenue, Cambridge CB3 0HE, UK\\
$^{8}$School of Chemical and Physical Sciences, Victoria University of Wellington, PO Box 600, Wellington 6140, NZ
}

\date{Accepted XXX. Received YYY; in original form ZZZ}

\pubyear{2019}

\volume{{\rm in press}}

\begin{document}
\label{firstpage}
\pagerange{\pageref{firstpage}--\pageref{lastpage}}
\maketitle

\begin{abstract}
We present the results of an intensive multi-epoch radio frequency campaign on the energetic and nearby \grb\ with the Karl G. Janksy Very Large Array and Arcminute Microkelvin Imager Large Array. We began observing \grb\ a day after its initial detection, and were able to monitor the temporal and spectral evolution of the source over the following weeks. The spectra and their evolution are compared to the canonical theories for broadband GRB afterglows, with which we find a general agreement. There are, however, a number of features that are challenging to explain with a simple forward shock model, and we discuss possible reasons for these discrepancies. This includes the consideration of the existence of a reverse shock component, potential microphysical parameter evolution and the effect of scintillation.
\end{abstract}




\section{Introduction}

Long gamma-ray bursts (GRBs) and their associated afterglow are some of the most energetic transient events in the Universe. They are believed to be the result of the core collapse of massive stars (see e.g. \citealt{levan2016} for a review of GRB progenitor systems) due in part to their association with supernovae (e.g. \citealt{galama2000}; \citealt{hjorth2003}), although supernova are not always found with long GRBs (e.g. \citealt{dellavalle2006}; \citealt{fynbo2006}; \citealt{hjorth2012}). The emission from GRBs is typically separated into two temporal phases: `prompt' emission in $\gamma$-rays, and `afterglow' emission which is broadband (ranging from radio to high energy X-ray frequencies) and can be detected months or even years after the GRB is first detected (e.g. \citealt{frail2000a}; \citealt{vanderhorst2008}). The prompt emission is thought to be produced in processes internal to the out-flowing material, with magnetic reconnection and internal shocks both invoked as the physical process responsible (\citealt{rees1994}; \citealt{spruit2001}; \citealt{granot2016}). The afterglow is the result of the ejected material interacting with the circumburst environment, which results in broadband synchrotron emission from shock accelerated electrons (\citealt{wijers1997}; \citealt{wijers1999}; \citealt{sari1998}). This outflow is believed to be an initially collimated and highly relativistic jet (\citealt{rhoads1997}; \citealt{rhoads1999}).\\

GRB afterglows are typically interpreted in the context of the fireball model (\citealt{meszaros1993}; \citealt{meszaros1997}; \citealt{piran1999}), detailing the interaction of the jet with the circumburst material. As the interaction occurs, two shocks form -- a forward shock (FS) and reverse shock (RS) -- and accelerate electrons into a power law energy distribution ($N(\gamma_\textrm{e})\propto\gamma_{\textrm{e}}^{-p}$, with $p$ typically in the range $2$ to $3$). These electrons emit via the synchrotron process as they spiral in the shock-enhanced magnetic field. The broadband spectrum from each shock is expected to be described by a series of power laws, with breaks occurring at the synchrotron self-absorption, minimum electron, and cooling frequencies ($\nu_{\textrm{a}}$, $\nu_{\textrm{min}}$ and $\nu_{\textrm{c}}$, respectively; \citealt{sari1998}). The exact values of the power law indices describing the spectrum depends on the ordering of these frequencies (\citealt{piran1999}), with $\nu_{\textrm{m}}<\nu_{\textrm{c}}$ defined as the slow cooling and $\nu_{\textrm{c}}<\nu_{\textrm{m}}$ the fast cooling case. The FS propagates into the circumburst material and is decelerated from its initially relativistic velocity. The temporal and spectral evolution of the emission from the FS depends on the density profile of the circumburst material, which is typically assumed to vary as $\rho(r)\propto r^{-k}$, with $k=0$ for a constant density (ISM-like; \citealt{sari1998}; \citealt{granot2002}) environment and $k=2$ for a wind-like environment, such as might be produced around a Wolf-Rayet star (e.g. \citealt{chevalier2000}; \citealt{vanerten2009}). The RS propagates \textit{back} into the ejected material and its final velocity depends on the velocity dispersion of the ejecta. In the Newtonian (or thin-shell) case the ejecta have a narrow distribution of Lorentz factors and as such the RS does not have a chance to become relativistic. In the relativistic, or thick-shell, case the dispersion of Lorentz factors is large enough such that the RS reaches relativistic velocities. The RS is short lived, and dominates at early times. Flashes of optical emission (e.g. \citealt{kulkarni1999opt}; \citealt{akerlof1999}) are an expected -- and observed -- feature of a reverse shock, as well as a radio flare which peaks on time scales of the order of days (e.g. \citealt{kulkarni1999rad}; \citealt{berger2003GRB020405}; \citealt{frail2000}; \citealt{anderson2014}). The FS contributes to the afterglow emission at all times. If the RS is relativistic, like with the FS, its evolution will be determined by the circumburst density profile (\citealt{kobayashi2000}; \citealt{zou2005}) whereas in the Newtonian case the radial Lorentz factor distribution of the burst material (with $\Gamma(r)\propto r^{-g}$ the often assumed form; e.g. \citealt{meszaros1999}) is the most important parameter governing the evolution (\citealt{kobayashi2000}; \citealt{chevalier2000}; \citealt{yi2013}).\\

Here we report on an extensive multi-frequency radio observing campaign on the nearby and energetic \grb, carried out with the Karl G. Jansky Very Large Array (VLA) and the Arcminute Microkelvin Imager Large Array (AMI-LA; \citealt{zwartAMI2008}; \citealt{hickishAMI2017}). We obtained a high cadence (22 observations made approximately every $1$-$4\,\textrm{d}$) light curve at $15.5\,\textrm{GHz}$ and multiple -- broadband -- VLA spectra of the source. These observations, combined with publicly available data at other wavelengths, allow us to track the spectral and temporal evolution of the source from early to late times and compare our observations with the theoretical understanding of GRB afterglows. This provides insight into the environment surrounding this energetic and nearby GRB.\\

\grb\ was detected at $T_{0}=\textrm{MJD}\ 58036.79$ by the Fermi Large Area Telescope (LAT) after an automated slew was triggered by the Fermi Gamma-ray Burst Monitor (GBM; \citealt{atwood2009}; \citealt{meegan2009}; \citealt{171010Afermigbm}; \citealt{171010Afermilat}). The GBM location was found to be consistent with the LAT position, measured as RA (J2000) = $66\fdg74$, Dec (J2000) = $-10\fdg53$ with a circular 90 per cent confidence region of radius $0.2\degr$, later refined to RA (J2000) = $66\fdg58$, Dec (J2000) = $-10\fdg46$ with a circular 90 per cent confidence region of radius $1.4$ arcsec from observations (\citealt{171010Aswift1}; \citealt{171010Aswift2}) obtained with the Neil Gehrels \textit{Swift} Observatory X-ray Telescope (XRT; \citealt{gehrelsSWIFT2004}). Optical follow-up showed that the position of the afterglow was consistent with a galaxy at redshift $z=0.33$, which was identified as the probable host (\citealt{171010Athorstensen}; \citealt{171010Akankare}). Considering a cosmological model (\citealt{spergel2003}) with $H_{0}=72\,\textrm{km}\,\textrm{s}^{-1}\textrm{Mpc}^{-1}$, $\Omega_{\textrm{M}}=0.27$ and $\Omega_{\Lambda}=0.73$ provides a luminosity distance to \grb\ of $1698\,\textrm{Mpc}$ for $z=0.33$ (\citealt{wright2006}). The $\gamma$-ray fluence of \grb, as measured by the Fermi-GBM (\citealt{171010Afermilat}), was $(6.42\pm0.02)\times10^{-4}\,\textrm{erg}\,\textrm{cm}^{-2}$ in the $10{-}1000\,\text{keV}$ energy band. Combining the distance with the fluence measured by the Fermi-GBM gives an isotropic equivalent gamma ray energy of $E_{\gamma}\sim2.2\times10^{53}\,\textrm{erg}$ in the rest frame of the source. \grb\ is therefore one of the most energetic long GRBs observed below a redshift of 0.5 (with an isotropic $\gamma$-ray energy release similar to GRBs 030329, 090818 and 130427A). It has been suggested (e.g. \citealt{guetta2007rates}) that while GRBs cover a wide range of luminosity space there are two distinct populations. These consist of common, under-luminous events only observable in the nearby universe and more energetic events which form the population observed between $z\sim2$ and $3$. These GRBs have isotropic $\gamma$-ray energies several orders of magnitude greater than those typically found at $z\lesssim0.5$ ($10^{52}{-}10^{54}\,\textrm{erg}$ verses $10^{48}{-}10^{52}\,\textrm{erg}$; see e.g. figure 1 in \citealt{perley2014}), and are rarely observed in the local universe due to both geometric and star formation rate considerations. These nearby yet luminous GRBs, such as GRB 171010A (along with the well studied GRBs 030329 and 130427A),  provide an excellent opportunity to probe the intricacies of afterglow emission.

\section{Observations}\label{sec:obs}

\subsection{Radio}\label{sec:radio_obs}
\subsubsection{Arcminute Microkelvin Imager Large Array}
As part of a GRB afterglow followup campaign (\citealt{anderson2018}) we began observing \grb\ using the AMI-LA at $T_{0}+1.27\,\text{d}$. An initial 4 hour observation was conducted at a central frequency of $15.5\,\textrm{GHz}$ with a $5\,\textrm{GHz}$ bandwidth spread over 4096 channels. Data were reduced using the Common Astronomical Software Applications (\textsc{casa}) package with a custom reduction pipeline, which flagged the data for radio frequency interference and instrumental effects, and calibrated the flux, bandpass and phases. We used 3C286 as the flux/bandpass calibrator, and J0438$-$0848 as the phase calibrator. Due to the low declination of \grb, the synthesised beam is elongated when compared to more northerly observations, with characteristic dimensions of $\sim100$ by $\sim35$ arcsec. We detected a bright ($3\,\textrm{mJy}$, with an image RMS of $\sim45\,\muup\text{Jy}$) unresolved source at phase centre (RA (J2000): $04^{\rm h}\,26^{\rm m}\,19\fs3\pm0\fs7$, Dec (J2000): $-10^{\degr}\,27\arcmin\,45.1\arcsec\pm10\arcsec$) in the initial image which overlaps with the \textit{Swift} 90 per cent confidence region and is $\sim7$ arcmin away from the nearest source in the NVSS catalogue (\citealt{condon1998}). We therefore identified the source as the radio afterglow of \grb. To calculate the source flux density we use the \textsc{casa} task \textsc{imfit}, adding a 5 per cent calibration error (in line with the RMS variability of the phase calibrator flux) in quadrature with the statistical one.\\

\begin{figure}
\includegraphics[width=\columnwidth]{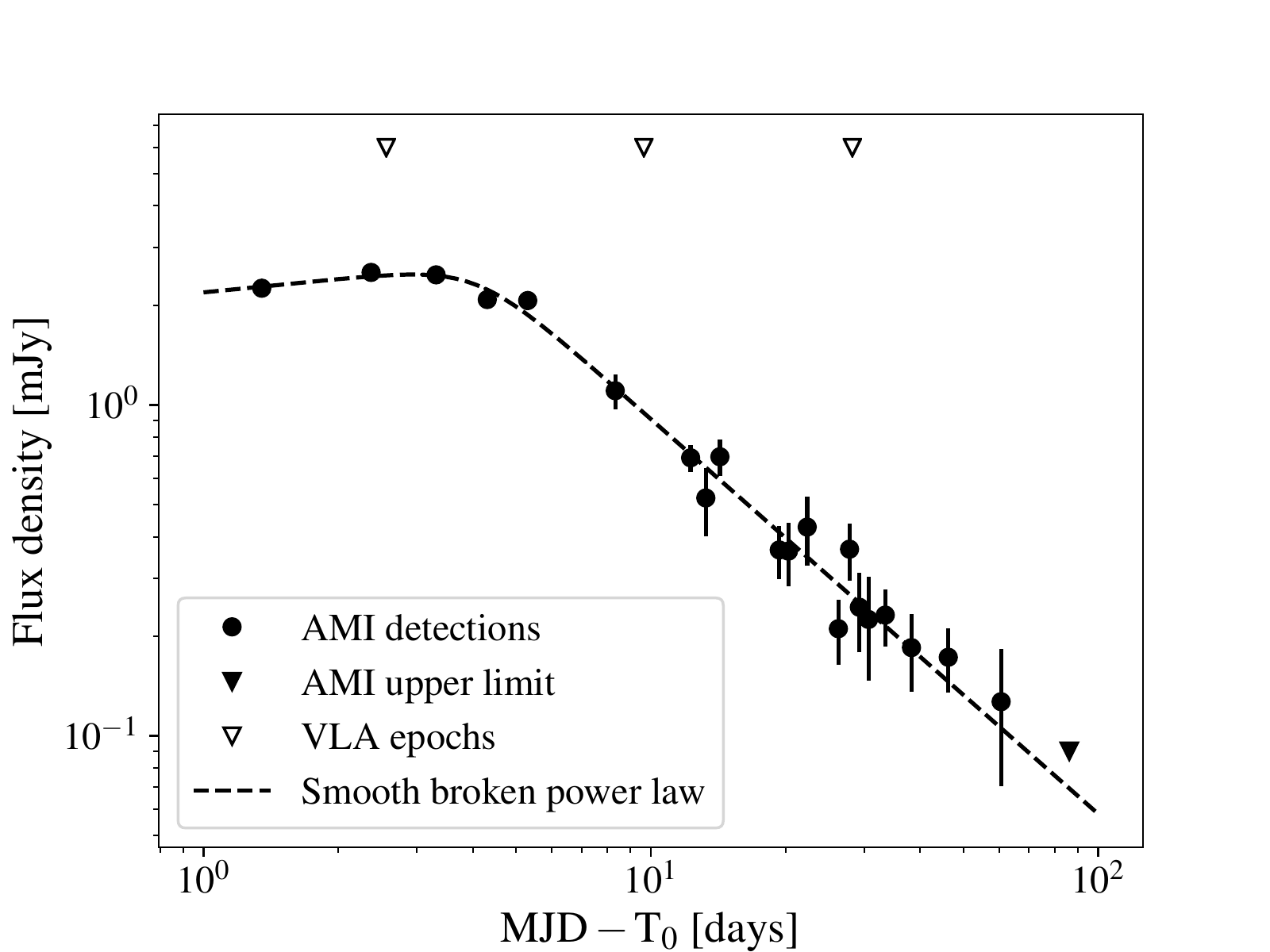}
\caption{$15.5\,\textrm{GHz}$ radio observations of \grb\ made with the AMI-LA. Filled circles and triangles represent AMI-LA detections and upper limits at a $3\sigma$ significance, respectively. The error bars demonstrate $1\sigma$ uncertainties. Unfilled triangles show the epochs at which VLA measurements were made. The dashed line is a smoothed broken power law fit to the data (excluding upper limits). The break time is found to be $T_{0}+4.1\pm0.4\,\textrm{d}$. The temporal power law index before and after the break is $\alpha=0.1\pm0.1$ and $\alpha=-1.19\pm0.06$, respectively. The reduced chi-square of the fit is $\chi^{2}_{\nu}=0.79$.}
 \label{fig:amilc}
\end{figure}

Upon detecting the radio afterglow of \grb\ we initiated a radio follow-up campaign which consisted of multiple epochs of AMI-LA observations, as well as multiple observations with the VLA. All other AMI-LA observations were carried out in the same instrumental configuration as the initial observation, and the reduction procedure was identical. A summary of the AMI-LA observing campaign is presented in Table \ref{table:amiobs} and the light curve is shown in Figure \ref{fig:amilc}.

\begin{table}
\centering
\caption{Summary of AMI-LA Observations of \grb. Errors are statistical and calibration (5\%) combined in quadrature. $\Delta T$ is the number of days between $T_{0}$ and the observation midpoint. The upper limit at $\Delta T=86.03\,\textrm{d}$ is three times the RMS noise in the image. All observations were taken at a central frequency of $15.5\,\text{GHz}$.}
\begin{threeparttable}
\label{table:amiobs}
\begin{tabular}{lccc}
\hline
Date & Obs. Length & Flux Density & Error\\
$[\Delta T]$ & [hours] & $[\textrm{mJy}]$ & $[\textrm{mJy}]$\\
\hline
1.35 & 4 & 2.26 & 0.13\\
2.37 & 3.5 & 2.52 & 0.14\\
3.31 & 4 & 2.48 & 0.14\\
4.31 & 4 & 2.08 & 0.13\\
5.31 & 4 & 2.07 & 0.14\\
8.33\tnote{a} & 3 & 1.11 & 0.13\\
12.27 & 4 & 0.69 & 0.07\\
13.27\tnote{a} & 4 & 0.52 & 0.12\\
14.27 & 4 & 0.70 & 0.09\\
19.37 & 3 & 0.37 & 0.07\\
20.30 & 4 & 0.36 & 0.08\\
22.36 & 3 & 0.43 & 0.10\\
26.25 & 4 & 0.21 & 0.05\\
27.80\tnote{b} & 8 & 0.37 & 0.07\\
29.24 & 4 & 0.25 & 0.07\\
30.66\tnote{b} & 8 & 0.23 & 0.08\\
33.42\tnote{b} & 8 & 0.23 & 0.05\\
38.24 & 4 & 0.19 & 0.05\\
46.21 & 4 & 0.17 & 0.04\\
60.66\tnote{a,b} & 11 & 0.13 & 0.06\\
86.15 & 4 & <0.09 & --\\
\hline
\end{tabular}
\begin{tablenotes}
\item [a] The source in these observations was not well fit by the clean beam due to residual noise. Thus the peak flux density was used.
\item [b] These observations combined data from multiple days in order to lower the noise level. In this case the time is the centroid of all observations used.
\end{tablenotes}
\end{threeparttable}
\end{table}

\subsubsection{Karl G. Jansky Very Large Array}
\begin{figure}
 \includegraphics[width=\columnwidth]{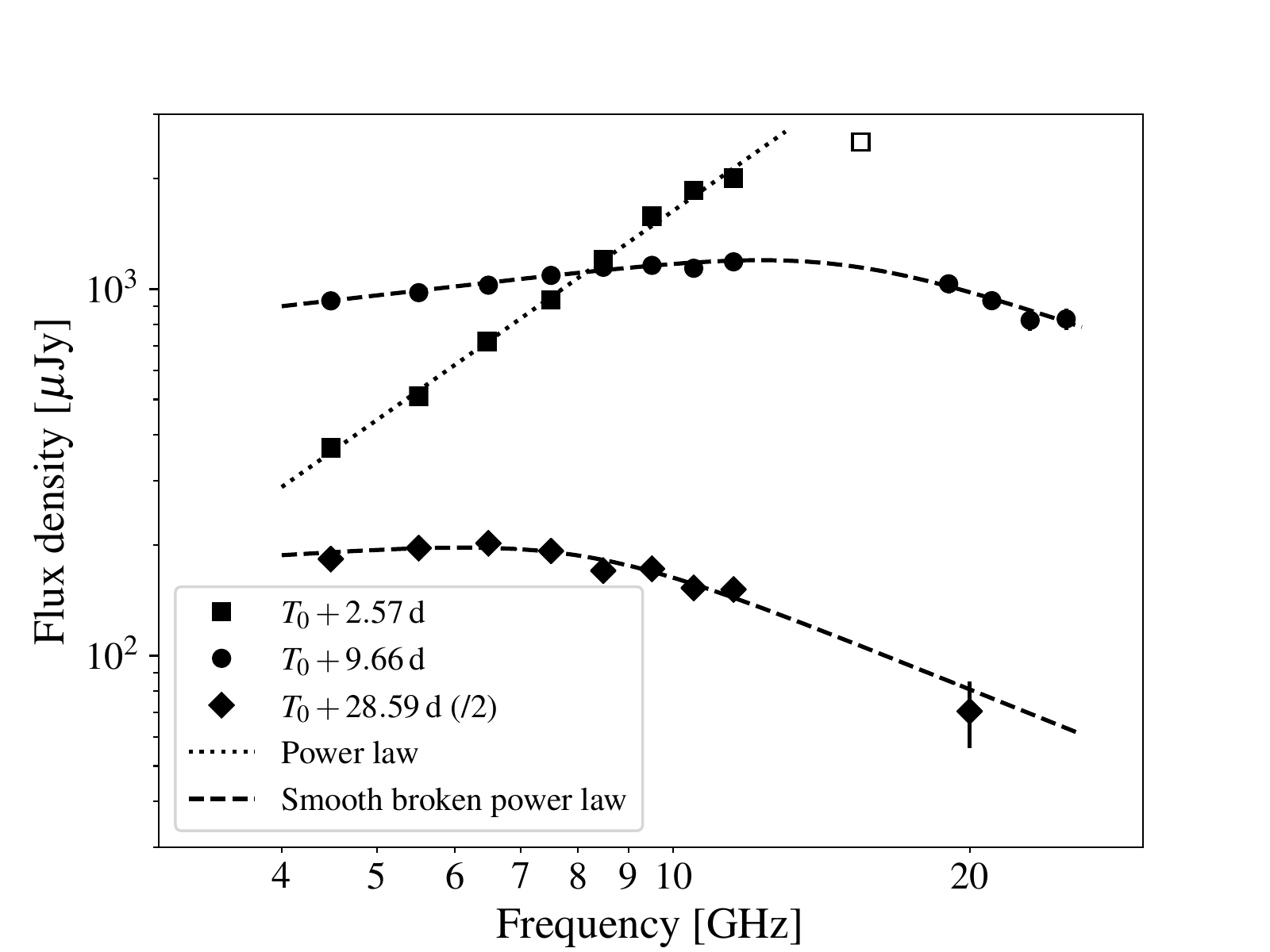}
 \caption{The radio spectrum of \grb\ between $4.5$ and $25\,\textrm{GHz}$ at three epochs. Squares, circles and diamonds are data from VLA observations made on $T_{0}+2.57\,\textrm{d}$, $T_{0}+9.66\,\textrm{d}$ and $T_{0}+28.59\,\textrm{d}$, respectively. The flux density measured from the third epoch has been divided by 2 for clarity. The error bars demonstrate $1\sigma$ uncertainties. The unfilled square is data from an AMI-LA observation which was taken $0.25\,\textrm{d}$ before the first VLA epoch. The dash-dot line shows a power law fit to the first epoch whereas dashed lines show smooth broken power law fits to the second and third epochs. The details of the fits are given in section \ref{sec:rad_sp} and are summarised in Table \ref{table:vlafits}. Data are given in Table \ref{table:vlaobs}.}
\label{fig:spec}
\end{figure}

We obtained three observations of \grb\ with the VLA (project code S81171) in B configuration at $2.57$, $9.66$ and $28.59\,\textrm{d}$ after the initial Fermi LAT detection. During the first epoch ($1\,\textrm{hr}$ total observation length) we recorded data between $\sim4$ and $\sim12\,\text{GHz}$ spread across 64 spectral windows, each consisting of 64 channels of width $2\,\textrm{MHz}$. During epochs two and three ($1.50\,\textrm{hr}$ each) we recorded data between $4$ and $12\,\text{GHz}$ and $18$ and $26\,\text{GHz}$, across a total of 128 spectral windows, each consisting of 64 channels of width $2\,\textrm{MHz}$ per channel. Data were calibrated in \textsc{casa} using the NRAO VLA scripted calibration pipeline with 3C138 and J0423$-$0120 as the absolute flux and phase calibrator, respectively. Imaging was also performed in \textsc{casa} using natural weighting, with a clean gain of $0.1$. In order to calculate spectral information across the VLA observing band we opt to split our data into $1-2\,\textrm{GHz}$ frequency chunks. To calculate the flux from the source (which is well detected as an unresolved source in all epochs) we use the \textsc{casa} task \textsc{imfit}. The only exception to this procedure was for the $18$ to $26\,\text{GHz}$ frequency range in the third epoch. At almost a month post detection, the source had become faint enough at these high frequencies such as to be no longer significantly detected in each $1$ or $2\,\textrm{GHz}$ range. To increase our sensitivity we therefore image using a $4\,\textrm{GHz}$ bandwidth between $18$ and $22\,\text{GHz}$ for this epoch. Data above $22\,\text{GHz}$ was not used due to artefacts corrupting the image in this frequency band. A summary of our VLA observations are given in Table \ref{table:vlaobs}, and the data are plotted in Figure \ref{fig:spec}. Note that while all frequencies within each epoch were not observed strictly simultaneously, they were observed $<1\textrm{hr}$ apart and as such we consider them simultaneous for the purpose of spectral fitting (and use the centroid time of each observation).

\begin{table}
\centering
\caption{Summary of VLA Observations of \grb. Errors are statistical and calibration (3\% for C and X band, 5\% for K band) combined in quadrature. $\Delta T$ is the number of days between $T_{0}$ and the observation midpoint.}
\begin{threeparttable}
\label{table:vlaobs}
\begin{tabular}{lcccc}
\hline
Date & Frequency & Bandwidth & Flux Density & Error\\
$[\Delta T]$ & $[\textrm{GHz}]$ & [\textrm{GHz}] & $[\muup\textrm{Jy}/\textrm{beam}]$ & $[\muup\textrm{Jy}/\textrm{beam}]$\\
\hline
2.57 & 4.5          & 1 & 369                & 21                      \\
2.57 & 5.5          & 1 & 510                & 21                       \\
2.57 & 6.5          & 1 & 719                & 26                       \\
2.57 & 7.5          & 1 & 936                & 33                       \\
2.56 & 8.5          & 1 & 1204               & 40                      \\
2.56 & 9.5          & 1 & 1580               & 54                       \\
2.56 & 10.5         & 1 & 1858               & 64                       \\
2.56 & 11.5         & 1 & 2009               & 71                       \\
9.66 & 4.5          & 1 & 929                & 42                       \\
9.66 & 5.5          & 1 & 978                & 33                       \\
9.66 & 6.5          & 1 & 1025               & 35                       \\
9.66 & 7.5          & 1 & 1090               & 35                       \\
9.67 & 8.5          & 1 & 1147               & 37                       \\
9.67 & 9.5          & 1 & 1161               & 38                       \\
9.67 & 10.5         & 1 & 1140               & 38                       \\
9.67 & 11.5         & 1 & 1187               & 41                       \\
9.64 & 19.1         & 2 & 1034               & 60                       \\
9.64 & 21.1         & 2 & 930                & 55                       \\
9.64 & 23.0         & 2 & 822                & 51                       \\
9.64 & 25.1         & 2 & 829                & 55                       \\
28.59 & 4.5         & 1  & 367                & 20                       \\
28.59 & 5.5         & 1  & 393                & 18                       \\
28.59 & 6.5         & 1  & 405                & 17                       \\
28.59 & 7.5         & 1  & 386                & 17                       \\
28.60 & 8.5         & 1  & 341                & 16                       \\
28.60 & 9.5         & 1  & 345                & 17                       \\
28.60 & 10.5        & 1  & 306                & 15                       \\
28.60 & 11.5        & 1  & 303                & 20                       \\
28.57\tnote{a} & 20.0        & 4    & 141                & 29\\                
\hline
\end{tabular}
\begin{tablenotes}
\item [a] This was a marginal detection, and so we assign it a conservative 20\% error.
\end{tablenotes}
\end{threeparttable}
\end{table}

\subsection{X-ray}
\begin{figure}
 \includegraphics[width=\columnwidth]{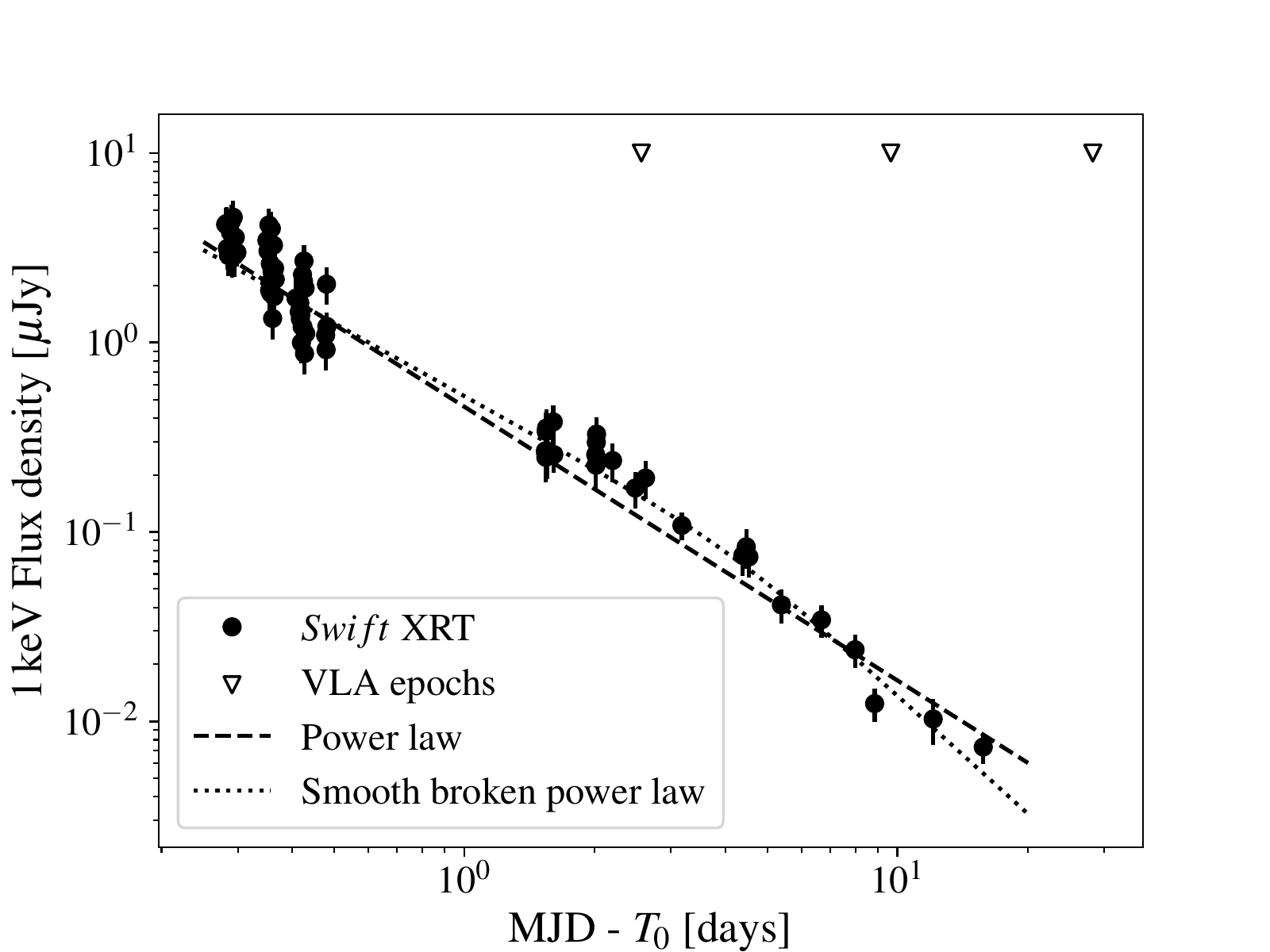}
 \caption{\textit{Swift}-XRT PC mode light curve for \grb. Solid points show the flux density at $1\,\textrm{keV}$. The dotted line shows a power law fit to the data with a temporal index of $\alpha=-1.45\pm0.03$. The reduced chi-square of the fit is $\chi^{2}_{\nu}=1.88$. The dotted line shows a smoothed broken power law fit to the data. The break occurs at $4\pm2\,\textrm{d}$ and the pre and post break slopes are $-1.18\pm0.05$ and $-2.1\pm0.3$, respectively. The reduced chi-square of the fit is $\chi^{2}_{\nu}=1.68$. Unfilled triangles show the epochs at which VLA measurements were made.}
 \label{fig:xrtlc}
\end{figure}

The afterglow of \grb\ was first observed in the $0.3{-}10\,\textrm{keV}$ energy band by the \textit{Swift} XRT at $T_{0}+6.72\,\textrm{hours}$, and then regularly over the next 16 days. \grb\ was observed for a total of $\sim104\,\textrm{ks}$ over 24 observing segments, with the majority of observations performed in photon counting (PC) mode. We retrieved the light curve and spectrum for \grb\ from the on-line burst analyser (\citealt{evans2007}; \citealt{evans2009}; \citealt{evans2010}). We use a photon index of 1.95 (from the best fit of the time averaged spectrum) when converting the observed flux to a flux density  at an energy of $1\,\textrm{keV}$ ($\nu=2.4\times10^{8}\,\textrm{GHz}$), following the method described in \citet{gehrels2008}. The \textit{Swift}-XRT light curve is shown in Figure \ref{fig:xrtlc}.


\section{Results}\label{sec:res}
Here we detail the results obtained from the X-ray and radio observations of \grb. We adopt the following convention when referring to the temporal and spectral flux dependence: $F(t,\nu)\propto t^{\alpha}\nu^{\beta}$. We refer to $\alpha$ as the temporal index and $\beta$ as the spectral index. We use separate subscripts for the radio and X-ray data to distinguish the indices, with subscripts R and X referring to the radio and X-ray properties respectively. When referring to the power law index of the time evolution of a break frequency we use $a$ ($\nu_{\textrm{break}}\propto t^{a}$), and the index of the flux evolution at a break frequency will be referred to as $b$ ($F_{\nu_{\textrm{break}}}\propto t^{b}$). 

\subsection{Radio spectra}\label{sec:rad_sp}
The spectra of \grb\ between $5$ and $25\,\textrm{GHz}$ at three different epochs are shown in Figure \ref{fig:spec}. Fitting an unbroken power law to the first epoch, we find a spectral index of $\beta_{\textrm{R}}=1.90\pm0.05$. Any break occurring at this time does so above approximately $10\,\textrm{GHz}$. We fit the second and third epochs (due to their obvious breaks) with smoothed broken power laws. We use a smoothing parameter of $s=5$ and the functional form presented in e.g. \citet{beuermann1999}. This defines a relatively sharp transition between power law segments, which we use for all smoothly broken power laws we fit in this work. For the second epoch we find that the spectrum breaks at $15^{+2}_{-2}\,\textrm{GHz}$, transitioning from $\beta_{\textrm{R}}=0.31^{+0.08}_{-0.06}$ to $\beta_{\textrm{R}}=-0.9^{+0.3}_{-0.3}$. For the third epoch we find a break at $8^{+1}_{-1}\,\textrm{GHz}$ transitioning from $\beta_{\textrm{R}}=0.2^{+0.3}_{-0.2}$ to $\beta_{\textrm{R}}=-1.1^{+0.2}_{-0.2}$. For epoch 1 errors and maximum likelihood values are derived from fitting using the python module \textsc{curve\_fit} (part of the \textsc{scipy} module; \citealt{scipy}). For epochs two and three we use a Markov Chain Monte Carlo sampler (\textsc{emcee}; \citealt{foremanmackey2013}) to fit the data. In this case the quoted best fit parameter value is the $50$\textsuperscript{th} percentile of the trace for a given parameter (after burn-in) while the lower and upper errors are the $16$\textsuperscript{th} and $84$\textsuperscript{th} percentile of the trace (after burn-in), respectively. The marginalised posterior distributions from this analysis are presented in the appendix, in Figures \ref{fig:vlaa1} and \ref{fig:vlaa2}. The marginalised posterior distributions are shown to consist of a single node (a single peak in the marginalised parameters spaces) for all parameter combinations. There is, however, some obvious parameter degeneracy for both epochs. In \ref{fig:vlaa3} we demonstrate the result of fitting both spectra simultaneously (with a broken power law), also using \textsc{emcee}. In doing so we fix the pre-break index to $1/3$, and make assumptions on the evolution of the break frequency and peak flux (details are given in Appendix A). While the results of this analysis do agree well with those derived from individually fitting the spectra, we do not use them in the main analysis presented in the paper. This is as we do not know a-priori if the break in each of the spectra is the same one, or which part of the broadband spectra we are sampling (assumptions which are made during the fitting presented in Appendix A for the joint spectral fit).

\begin{table*}
\centering
\caption{Summary of best fit parameters when fitting the VLA radio spectra of \grb. Details of the techniques used to calculate the best fit values and their errors are presented in Section \ref{sec:rad_sp}.}
\label{table:vlafits}
\begin{tabular}{ccccc}
\hline
VLA Epoch ($\Delta T$) & Pre break slope & Post break slope & Break frequency & $\chi^{2}_{\nu}$ \\
Days & -- & -- & $\text{GHz}$ & -- \\
\hline
2.57 & $1.90\pm0.05$ & -- & $\gtrsim10\,\text{GHz}$ & 1.43 \\[0.125cm]
9.66 & $0.31^{+0.08}_{-0.06}$ & $-0.9^{+0.3}_{-0.3}$ & $15^{+2}_{-2}$ & 0.34 \\[0.125cm]
28.59 & $0.2^{+0.3}_{-0.2}$ & $-1.1^{+0.2}_{-0.2}$ & $8^{+1}_{-1}$ & 0.84 \\
\hline
\end{tabular}
\end{table*}

\subsection{Radio light curve}
The AMI-LA $15.5\,\textrm{GHz}$ light curve (shown in Figure \ref{fig:amilc}) demonstrates almost no flux evolution between $1.27\,\textrm{d}$ (the first AMI-LA observation) and $4.1\pm0.4\,\textrm{d}$ post burst (the fitted time of the temporal break, $3.1\pm0.3\,\textrm{d}$ in the GRB rest frame). The break time is computed using a smoothed broken power law fit, as clearly the first three data points do not follow the decaying trend at later times. The pre-break index is $\alpha_{\textrm{R}}=0.1\pm0.1$ and the post-break index is $\alpha_{\textrm{R}}=-1.19\pm0.06$. At around $80\,\textrm{d}$ the flux density from \grb\ has reached the sensitivity limit of the AMI-LA and we are only able to place upper limits at $\sim90\,\muup\textrm{Jy}$ on the $15.5\,\textrm{GHz}$ emission which is not accounted for in the fitting process.

\subsection{X-ray light curve and spectrum}
The X-ray light curve for \grb\ (Figure \ref{fig:xrtlc}) shows the flux density declining from $\sim6$ to $\sim0.01\,\muup\textrm{Jy}$ over the 20 day observing period. Fitting a single power law decay to the data gives a temporal index of $\alpha_{\textrm{x}}=-1.45\pm0.03$. We also demonstrate a smoothed broken power law fit, with a break occurring at $4\pm2\,\textrm{d}$, and a pre and post break slope of $-1.18\pm0.05$ and $-2.1\pm0.3$, respectively. The time averaged (between $T_{0}+0.28\,\textrm{d}$ and $T_{0}+0.48\,\textrm{d}$) X-ray spectrum is well described ($\chi^{2}_{\nu}=0.87$) by an absorbed power law with a photon index of $\Gamma=2.0^{+0.2}_{-0.1}$ and a total column density of $N_{\textrm{H}}=3.4^{+0.7}_{-0.6}\times10^{21}\,\textrm{cm}^{-2}$ (\citealt{171010Aswift2}) which is in excess of the galactic value of $N_{\textrm{H}}\sim7\times10^{20}\,\textrm{cm}^{-2}$, indicating absorption intrinsic to the source or its host galaxy. The measured photon index, $\Gamma$, gives a spectral index of $\beta_{\textrm{X}}=\Gamma-1=1.0^{+0.2}_{-0.1}$. 

\section{Discussion}\label{sec:dis}
We will begin with attempting to explain our radio and X-ray observations in the context of the fireball model with a single FS component. This is a good first order step, as typically GRBs showing reverse shock radio emission are already declining on $\sim1\,\textrm{d}$ timescales (e.g. \citealt{berger2003GRB020405}, \citealt{frail2000}, \citealt{anderson2014}) whereas in our case the decline occurs a few days later. We then discuss issues with this simple model, and potential solutions.\\

\begin{table}
\centering
\caption{Summary of movement of characteristic frequencies of \grb.}
\label{table:breakmove}
\begin{tabular}{cccc}
\hline
& $\nu_{\textrm{a}}$ & $\nu_{\textrm{m}}$ & $\nu_{\textrm{c}}$ \\
\hline
$a$ & $\lesssim-0.7$ & $-0.6\pm0.2$ & unknown \\
\hline
\end{tabular}
\end{table}

\subsection{A Forward shock model}
\subsubsection{Radio}
All three VLA spectra are well described by simple single or broken power laws. The slope in the first epoch, $\beta_{\textrm{R}}=1.90\pm0.05$, is consistent with synchrotron self absorption and $10\,\text{GHz}\lesssim\nu_{\textrm{a}}<\nu_{\textrm{m}}$. In that case, the theoretical spectral index is $\beta=2$, while for a self-absorbed spectrum with $10\,\text{GHz}\lesssim\nu_{\textrm{m}}<\nu_{\textrm{a}}$ we would expect a steeper spectrum with an index of $\beta=2.5$. There is the hint of a turnover at the highest VLA frequency at the first epoch ($11.5\,\text{GHz}$), which is further supported by the closest in time AMI-LA observation taken $0.25 \,\textrm{d}$ earlier. However, given that the observation was not simultaneous and the spectrum is evolving most quickly at early times, we set the limit $\nu_{\textrm{a}}\gtrsim10\,\textrm{GHz}$. In the second and third epochs the self absorption break has moved below the VLA observing band, and the spectra are each well described by a broken power law with consistent indices below and above an evolving break frequency. A spectral slope of $\beta=1/3$ is expected between $\nu_{\textrm{a}}$ and $\nu_{\textrm{m}}(<\nu_{\textrm{c}})$, while the spectral slope between $\nu_{\textrm{m}}$ and $\nu_{\textrm{c}}$ is expected to be $\beta=-(p-1)/2$ (regardless of the density profile of the surrounding medium). Comparing these theoretical slopes with the observed ones suggests that the spectral break in both epochs is due to the peak frequency. The observed slopes below the break are consistent with $\beta=1/3$, and using the best-constrained spectral slope above the break (for epoch 3) we obtain $p=2.6^{+0.4}_{-0.4}$. The steepness of the post-break slope also confirms the order of the breaks to be $\nu_{\textrm{a}}<\nu_{\textrm{m}}<\nu_{\textrm{c}}$, as a shallower slope with $\beta=-0.5$ would be expected between the peak and cooling breaks if they were reversed. This is unsurprising since the cooling break is expected to reside at higher frequencies, and the minimum energy break moves to lower frequencies faster than the cooling break, regardless of the surrounding density profile.\\

The fireball model makes predictions for the temporal evolution of the spectral breaks, the flux at these breaks, and the flux in the different segments of the broadband spectrum. The evolution is tied to the density profile of the region surrounding the GRB. We see a very different spectrum between epochs one and two, with no sign of self absorption in epoch two. The self absorption break has therefore moved through and below our observing band, and we can constrain the self-absorption frequency to be $\nu_{\textrm{a}}\lesssim4.5\,\textrm{GHz}$ at the second epoch. We can therefore put a limit on the evolution of the self absorption break frequency, assuming that it is a power-law, to have an index of $a\lesssim-0.7$ ($\nu_{\rm sa}\propto t^{a}$). The limit placed on the temporal evolution of $\nu_{\textrm{a}}$ is most consistent with a steep density profile (wind-like) with $k\gtrsim2.2$ (see e.g. table 5 in \citealt{vanderhorst2014}). While the self-absorption frequency has moved through the band, a second and distinct spectral break is present in the second epoch at $15^{+2}_{-2}\,\textrm{GHz}$ and in the third epoch at $8^{+1}_{-1}\,\textrm{GHz}$. Identification of this break in the context of the forward shock model proves to be more puzzling, even though the spectral slopes below and above the break indicate that this is $\nu_{\textrm{m}}$. From the spectral fitting we derive a value for the time evolution of this break which, again assuming a power law, has an index $a=-0.6\pm0.2$ (this is constrained significantly better by jointly fitting the spectra, as shown in Appendix A). The movement of the break frequencies are summarised in Table \ref{table:breakmove}. These results from individually fitting epochs two and three are confirmed when jointly fitting the spectra as detailed in Figure \ref{fig:vlaa3} and discussed in the appendix. The prediction is that $\nu_{\textrm{m}}$ should, for all density profiles, move towards lower frequencies with a temporal power law index of $-1.5$. While we do see this break moving in the expected direction, it is moving significantly slower. The flux evolution at the break, which evolves with $b\sim-1$, is consistent with the minimum energy break for $k\sim2.7$ (or a post jet break spectral peak; \citealt{vanderhorst2005}; \citealt{vanderhorst2014}). The presence of a jet break at $\sim4\,\textrm{d}$ (as hinted at by fitting the X-ray light curve) would cause $\nu_{\textrm{m}}$ to move to lower frequencies even faster and so we disfavour it.\\

Considering the temporal flux evolution presents additional complications for a simple forward shock model. From the AMI-LA ($15.5\,\textrm{GHz}$) light curve, we see a clear break at around $4\,\textrm{d}$, which falls between the first and second VLA epochs. It is clear that the spectral break we see in the second and third epochs is not responsible for the temporal break we see with the AMI-LA, as if we extrapolate back in time at its derived rate it would be found at around $25\,\textrm{GHz}$ at $4.1\,\text{d}$ (the observed VLA break has only just reached $15\,\text{GHz}$ at $10\,\text{d}$ post burst so it is unlikely it caused the AMI-LA turnover). Therefore the self-absorption break is the most likely cause of the break in the AMI-LA band. When considering the AMI-LA light curve before its temporal break (which we infer is due to the self-absorption break, and thus in the range $\nu<\nu_{\textrm{a}}<\nu_{\textrm{m}}<\nu_{\textrm{c}}$) we see an almost constant flux. This is hard to explain for any density profile considering just a forward shock (see table 5 in \citealt{vanderhorst2014}). After the turnover in the AMI light curve (caused by the passage of the self absorption break through the AMI-LA band) the decay is well described by a single power-law, despite the fact that what we infer as the minimum energy break must also move through the AMI-LA band before $\sim10\,\text{d}$. Again referring to table 5 in \citet{vanderhorst2014}, it is possible for the flux evolution to be the same on either side of the minimum energy break. In the region $\nu_{\textrm{a}}<\nu<\nu_{\textrm{m}}<\nu_{\textrm{c}}$ (observed with the AMI-LA at $T_{0}+4.1$ to $\sim T_{0}+10\,\text{d}$) the decay slope can be explained with a steep density profile with $k\sim3.1$ which is broadly consistent with estimates for the density profile we have derived previously. However, the minimum energy break must move through the AMI-LA band, but we see no change in the temporal slope. To remain consistent with this requires an extremely hard electron energy index (which contradicts the measured VLA spectral index) of $p\sim0.6$. 

\subsubsection{X-ray}
Given that the forward shock model predicts a simple spectral shape across a huge (radio through to X-ray) range of frequencies, we can use the \textit{Swift} data to further constrain the parameters for \grb. We can attempt to identify the section on the spectrum that the $1\,\text{keV}$ emission is from, and thus where it lies with respect to the characteristic frequencies. The X-ray spectrum is well described by a power law with index $\beta_{X}=1.0^{+0.2}_{-0.1}$, which is similar to the spectral slope above the break as measured by the VLA. It is unclear, from just measuring the spectral slope, from which section of the spectrum the $1\,\text{keV}$ emission is from. Looking at the light curve, which decays with $\alpha_{\text{X}}=1.45\pm0.03$ (considering only the power law fit) we immediately confirm that the X-ray emission is above the minimum energy break, as we only see the flux density decline. The GRB closure relations (e.g. \citealt{racusin2009}) favour $\nu_{\textrm{c}}<\nu_{\textrm{X}}$, but given the error on the X-ray spectral slope it is far from definitive. For the broken power law fit, the decline post break is too steep to be explained by the cooling frequency moving through the \textit{Swift} observing band. The break is, however, consistent in time with the break at $15.5\,\text{GHz}$. Such an achromatic break, if real, could indicate that a jet break occurred at around this time, although it is difficult to confirm the legitimacy of this feature.\\

To summarise, we find that the most likely explanation of the data in the context of the forward shock model is as follows. At early times ($<5\,\textrm{d}$) we have that $15.5\,\textrm{GHz}<\nu_{\textrm{a}}<\nu_{\textrm{m}}<\nu_{\textrm{c}}<\nu_{\textrm{X}}$. Then at $\sim5\,\textrm{d}$ the self absorption break moves through the AMI-LA band and $\nu_{\textrm{a}}\lesssim15.5\,\textrm{GHz}<\nu_{\textrm{m}}<\nu_{\textrm{c}}<\nu_{\textrm{X}}$ with our VLA measurements sampling the minimum electron energy break at the second and third epochs. The location of the X-ray emission is not well constrained. We find the data to be most consistent with a steep density profile, with $k\sim3$, although there are significant deviations from a simple forward shock model.

\subsection{Inconsistencies: Explaining the unusual behaviour}
While we have attempted to contextualise our observations in terms of the forward shock model, there are obvious inconsistencies with the picture we have provided. For example, we would expect to see two breaks in the AMI-LA light curve when $\nu_{\textrm{a}}$ and then $\nu_{\textrm{m}}$ move through the band. Also, the movement of the $\nu_{\textrm{m}}$ is significantly slower than that predicted by a simple forward shock model. As we have inferred the movement of the self absorption break below all of our observing bands, there are only two possible orderings of the minimum energy and cooling frequencies. Previously we assumed the most likely configuration and the location of the radio band in this configuration, but it is also worth considering if this could be the cooling break moving through the radio band. For both the fast and slow cooling cases this is easy to rule out. For fast cooling, there is no good value of $k$ for which the cooling break moves to lower frequencies and the flux drops off as quickly as we observe between epochs 2 and 3. For the fast cooling case the spectrum should have a negative index before and after the break, whereas we clearly see a rise before the break and so rule this out. Below we detail some possible explanations for the puzzling behaviour of \grb.

\subsubsection{A reverse shock component}
\begin{figure}
 \includegraphics[width=\columnwidth]{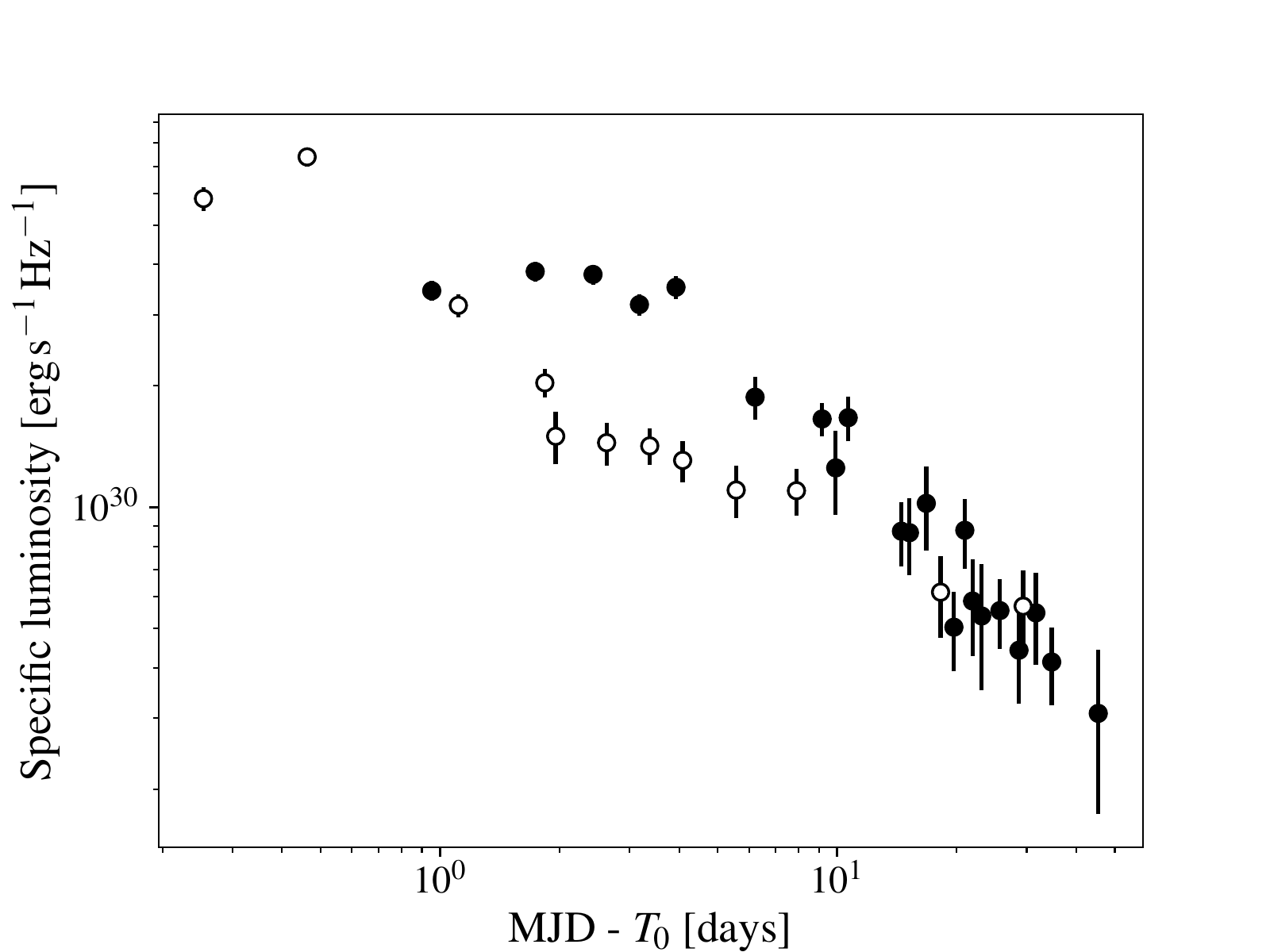}
 \caption{Specific luminosity of GRB 130427A (unfilled circles) and GRB 171010A (filled circles) at $15.7\,\text{GHz}$ and $15.5\,\text{GHz}$, respectively, made with the AMI-LA. We have k-corrected each section of both light curves according to \citet{chandra2012}. For GRB 171010A the spectral and temporal indices used for the correction were taken from this work. For GRB 130427A they were taken from \citet{perley2014} and \citet{anderson2014}.}
 \label{fig:13vs18}
\end{figure}

Perhaps most perplexing is the apparent movement of the break frequency between VLA epochs two and three. We established above that this is most likely the peak frequency $\nu_{\textrm{m}}$. We measure $\nu_{\textrm{m}}$ moving with a power law index $a=-0.6\pm0.2$, as opposed to the theoretical expectation of $a=-1.5$. In the context of a simple forward shock this discrepancy is very hard to explain. We briefly consider the effect of including emission from a reverse shock in our model for \grb. To properly constrain forward + reverse shock models, it is vital to have observational data across the electromagnetic spectrum from early times, particularly at optical, sub-mm and radio frequencies where the reverse shock will dominate. Unfortunately GRB 171010A was not observed extensively at all of these frequencies (from an early time) and so constraining the properties of a reverse shock is quite difficult. We can, however, make some qualitative statements on the effect of a reverse shock based on more completely observed GRBs. GRB 130427A is an obvious source to compare with. It is at a very similar redshift to \grb\ and also had an isotropic energy release more typical of GRBs at larger redshifts, just like \grb. The AMI-LA carried out long term monitoring on GRB 130427A, and a comparison to \grb\ is shown in Figure \ref{fig:13vs18}. A reverse shock component was invoked (e.g. \citealt{laskar2013}; \citealt{vanderhorst2014}; \citealt{anderson2014}; \citealt{perley2014}) to explain both the AMI-LA light curve as well as the broadband spectrum of GRB 130427A. Both GRBs show a slowly rising/flat early time $15.5\,\text{GHz}$ light curve, which turns over and decays after a few days. GRB 130427A, however, turns over close to $1\,\text{d}$ after the initial detection, faster than for \grb. This timescale is a clear sign, along with the additional change in slope at around 4 days, of a reverse shock dominating the radio emission at early times and becoming sub-dominant later on. No such secondary break is seen for GRB 171010A. \citet{perley2014} attributed the early peak in GRB 130427A's radio light curve as being due to the self-absorption frequency of the reverse shock (although it was later shown by \citealt{vanderhorst2014} that $\nu_{\textrm{m}}$ and $\nu_{\textrm{a}}$ for a thin-shell reverse shock model were at similar frequencies at the time of this peak). Our analysis of \grb\ determines that it is likely a self absorption break that causes the early peak in \grb. If this break had come from a reverse shock we would perhaps expect to see a secondary break in the light curve as was seen with GRB 130427A. If, however, the reverse shock dominates until a few days, we would predict that the secondary break may also occur later and so this secondary turnover may have happened after \grb\ was no longer detected by the AMI-LA. However, we could just be seeing the transition from reverse shock to forward shock domination and have missed the RS peak, which would be the natural assumption if not for the unusual spectral behaviour. \citet{alexander2017b} also invoked a reverse shock to explain the early time radio emission from GRB 160625B, whereas \citet{laskar2018} invoked two reverse shock components (one caused by a shell collision) to explain the broadband emission properties of GRB 140304A. Our relatively limited spectral coverage, as well as small number of spectral epochs, makes confirming the presence of a reverse shock difficult. However the relatively early radio peak, identified as being due to a self-absorption break, along with the unusual spectral evolution are compelling puzzles. This is especially true as the observed movement of $\nu_{\textrm{m}}$ in GRB 171010A of $a=-0.6\pm0.2$ is not consistent with the predicted movement for a reverse shock alone (for any density profile), and thus we would require two spectral components to explain the movement of this break. This again is challenging for the simple spectral shapes we observe. It is, however, potentially unsurprising that a basic reverse shock model does not completely describe the data. The simple density structure usually invoked for the ejected material (inside which the reverse shock occurs) likely does not accurately reflect physical conditions in the jet. More complex density structures in the out-flowing material could lead to significantly more complex overall evolution, most notably early time X-ray flares (e.g. \citealt{hascoet2017}).

\subsubsection{Evolution of microphysical parameters}
It has been suggested (e.g. \citealt{filgas2011}; \citealt{vanderhorst2014}) that time evolution of microphysical parameters such as the fraction of energy in electrons and magnetic field ($\epsilon_{e}$ and $\epsilon_{B}$) could account for the deviation of $\nu_{\textrm{m}}$ from the predicted value. For $p>2$ and adiabatic evolution it can be shown (e.g. \citealt{piran1999}; \citealt{panaitescu2000}) that  $\nu_{\textrm{c}}\propto\epsilon_{B}^{-3/2}t^{-1/2}$ and $\nu_{\textrm{m}}\propto\epsilon_{B}^{1/2}\epsilon_{e}^{1/2}t^{-3/2}$. As we have no constraint on the evolution of the movement on $\nu_{\textrm{c}}$ we can state generally that if both $\epsilon_{e}$ and $\epsilon_{B}$ are functions of time as $\epsilon_{e}\propto t^{\xi}$ and $\epsilon_{B}\propto t^{\psi}$ then $\xi+\psi\approx1.8$ to match our observations. The required evolution is therefore not modest, even if it is distributed approximately evenly between the electrons and the magnetic field. Even more problematic is that one expects, based on theoretical considerations and simulations, $\epsilon_{e}$ and $\epsilon_{B}$ to decline over time, not increase. We therefore strongly disfavour microphysical parameter evolution as the primary driver of the unusual spectral/temporal evolution.\\

\subsubsection{Scintillation effects}
It is also worth considering the possible effect of scintillation on our radio observations of \grb. Scintillation is the distortion of electromagnetic radiation as it moves through the ISM, and is more pronounced for radiation from compact sources, such as GRBs. For example, observations of GRBs 160625B and 161219B show scintillation inducing large amplitude variability in flux and radio spectral index on timescales as short as minutes, as well as unusual spectral shapes, across the radio band (\citealt{alexander2017b}; \citealt{laskar2018b}; \citealt{alexander2018}). To check for short time-scale scintillation effects in our radio data we imaged each VLA epoch in $5\,\textrm{m}$ time chunks in each of the frequency bands used to create the spectra shown in Figure \ref{fig:spec}. We see no evidence of significant short term variability in any of the bands. Scintillation can also occur on longer timescales, but given our simple spectra (well described by a single or broken power-law) with similar slopes across epochs 2 and 3, we disfavour scintillation causing the unusual spectral break evolution.\\

Other more complex pictures have also been invoked to explain unusual GRB behaviours. These include multiple jet components (e.g. \citealt{berger2003GRB030329}; \citealt{vanderhorst2014}) as well as variable energy injection into the fireball through either external density variations in the surrounding material (e.g. \citealt{lazzati2002}; \citealt{heyl2003}) or refreshed shocks from variable Lorentz factors in the out-flowing material (e.g. \citealt{rees1998}; \citealt{guetta2007GRB050713A}). While these additional complications are worth considering, they require broadband observations (particularly early-time optical data), which are not presented in this work.  When a more complete set of observations for \grb\ is available these models will be worth further analysis.

\section{Conclusions}\label{sec:con}
This work highlights the benefits of early and regular monitoring of long GRBs at multiple radio frequencies, as well as in other wavebands. Having high cadence light curves and multiple broadband spectral energy distributions, is essential to characterise the spectral and temporal behaviour of this source class. In some ways \grb\ can be aptly described by a simple forward shock model, propagating into a medium with a steep density profile. However, the evolution of the spectra and a number of features of the light curves are hard to explain. Most striking is the extremely slow movement of the peak frequency, which should evolve to lower frequencies faster than observed. We have considered several possibilities for this discrepancy, including time evolution of microphysical parameters, extreme scintillation and the addition of a reverse shock but cannot conclusively attribute any of these to the observed unusual evolution.

\section*{Acknowledgements}
We thank the anonymous referee for their helpful comments which helped to improve this work. JSB would like to acknowledge the support given by the Science and Technology Facilities Council through an STFC studentship. A.H. acknowledges support by the I-Core Program of the Planning and Budgeting Committee and the Israel Science Foundation. GEA is the recipient of an Australian Research Council Discovery Early Career Researcher Award (project number DE180100346) funded by the Australian Government. We thank the Mullard Radio Astronomy Observatory staff for scheduling and carrying out the AMI-LA observations. The AMI telescope is supported by the European Research Council under grant ERC-2012-StG-307215 LODESTONE, the UK Science and Technology Facilities Council, and the University of Cambridge. This work made use of data supplied by the UK Swift Science Data Centre at the University of Leicester. The National Radio Astronomy Observatory is a facility of the National Science Foundation operated under cooperative agreement by Associated Universities, Inc.




\bibliographystyle{mnras}
\bibliography{paper_library} 



\newpage
\appendix

\section{\textsc{emcee} spectral fitting}

\begin{figure}
 \includegraphics[width=\columnwidth]{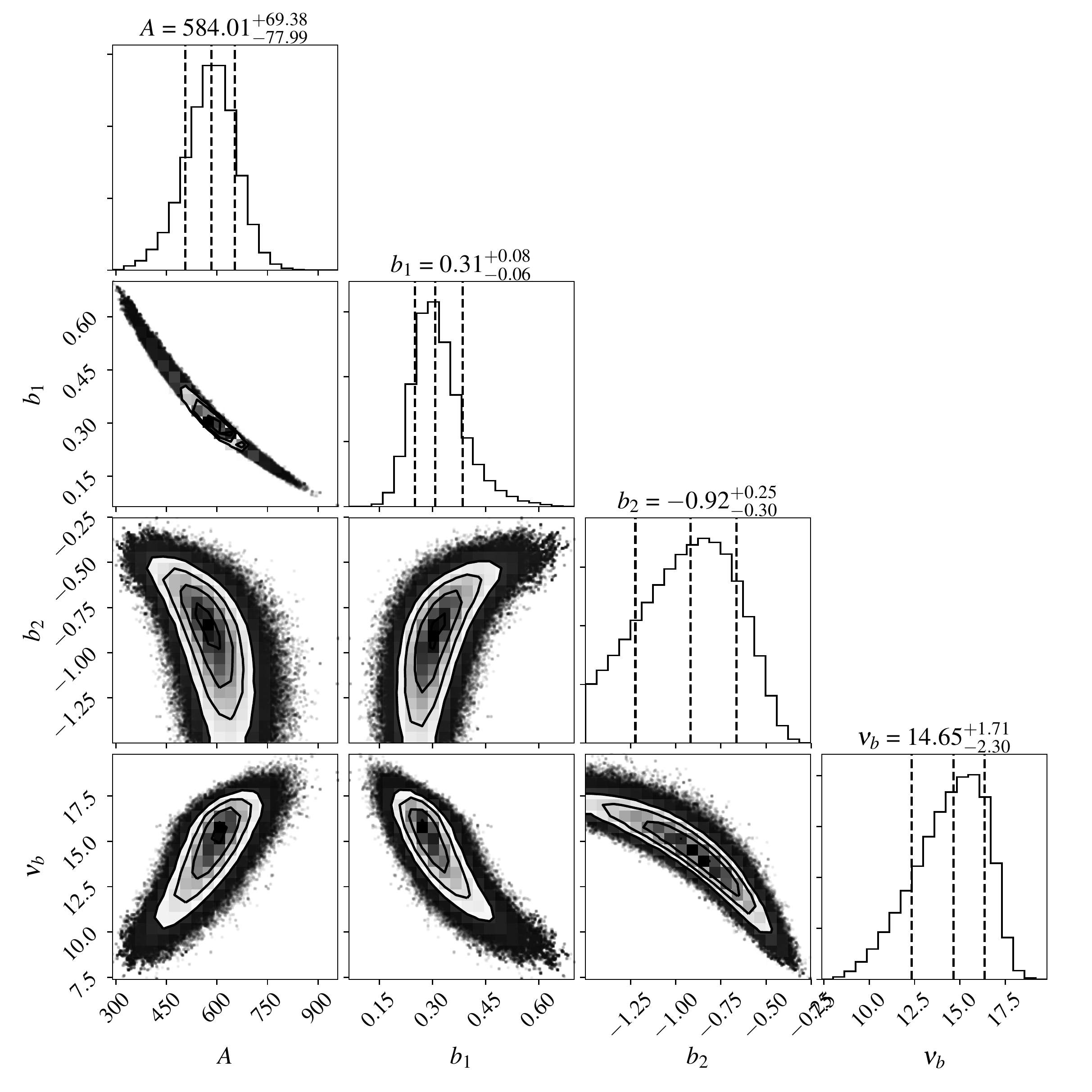}
 \caption{Marginalised posterior distribution for the smooth broken power law fit of the VLA epoch 2 spectrum. $A$, $b_1$, $b_2$, $\nu_b$ are the maximum flux, pre break slope, post break slope and break frequency respectively. The dotted lines indicate the $16^{\rm th}$, $50^{\rm th}$ and $84^{\rm th}$ percentiles on the parameters which are used to estimate the error.}
 \label{fig:vlaa1}
\end{figure}

\begin{figure}
 \includegraphics[width=\columnwidth]{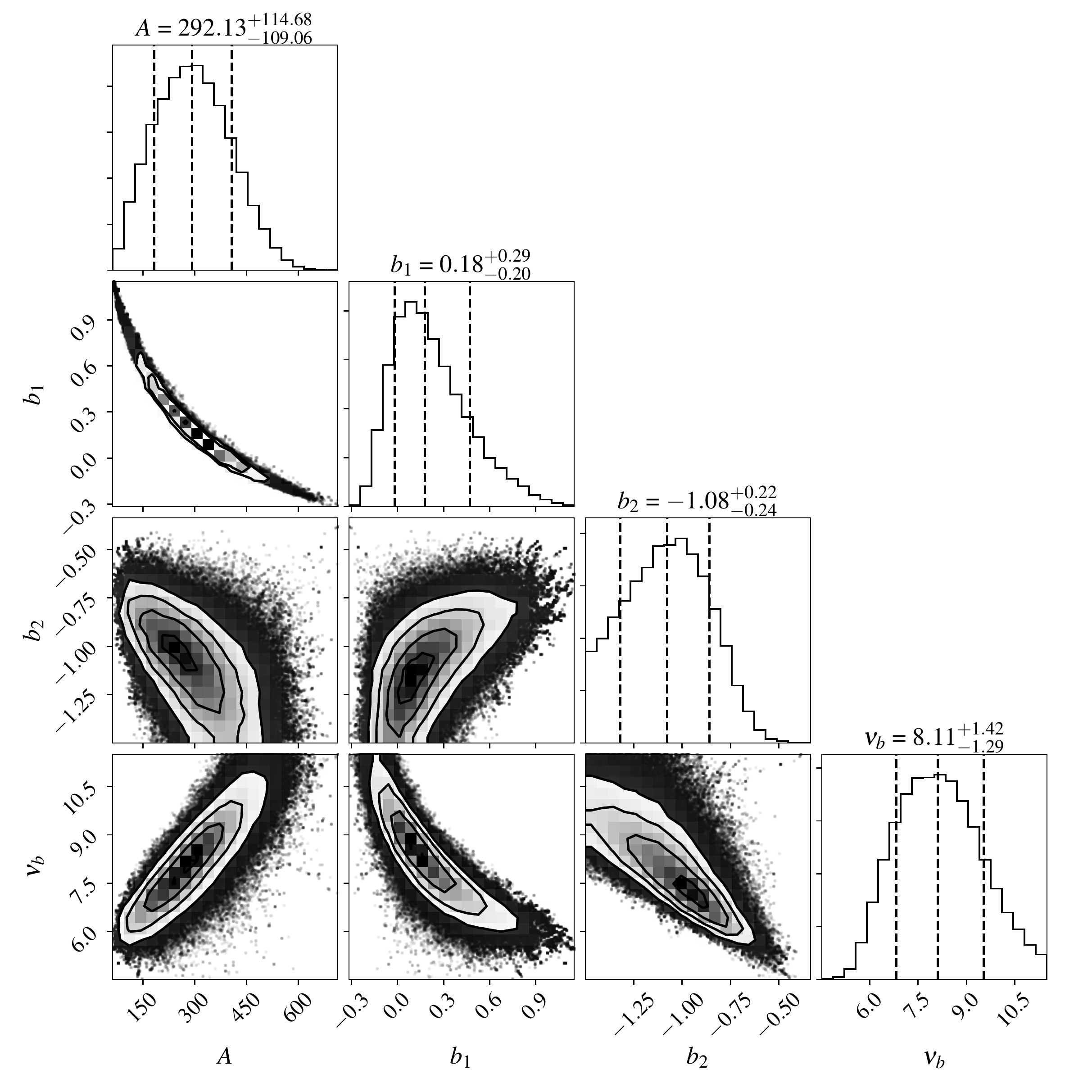}
 \caption{Marginalised posterior distribution for the smooth broken power law fit of the VLA epoch 3 spectrum. $A$, $b_1$, $b_1$ and $\nu_b$ are the maximum flux, pre break slope, post break slope and break frequency respectively. The dotted lines indicate the $16^{\rm th}$, $50^{\rm th}$ and $84^{\rm th}$ percentiles on the parameters which are used to estimate the error.}
 \label{fig:vlaa2}
\end{figure}

Here we present additional information on the spectral fits discussed in the main body of this work. Figures \ref{fig:vlaa1} and \ref{fig:vlaa2} show the marginalised posterior distributions for the second and third epochs VLA spectra when fit with a smooth broken power law. Uniform priors were used on all parameters, and we limit the post break spectral index such that $b_{2}>-1.5$. This corresponds to the steepest theoretically predicted slope (fast cooling) for $p=3$. We describe a smooth broken power law as $F(\nu)=A\nu_{b}^{b1}((\nu/\nu_{b})^{-b_{1}s} + (\nu/\nu_{b})^{-b_{2}s})^{-1/s}$, as described in \citet{beuermann1999} and \citet{schulze2011}, and we fix $s=5$. 

\begin{figure}
 \includegraphics[width=\columnwidth]{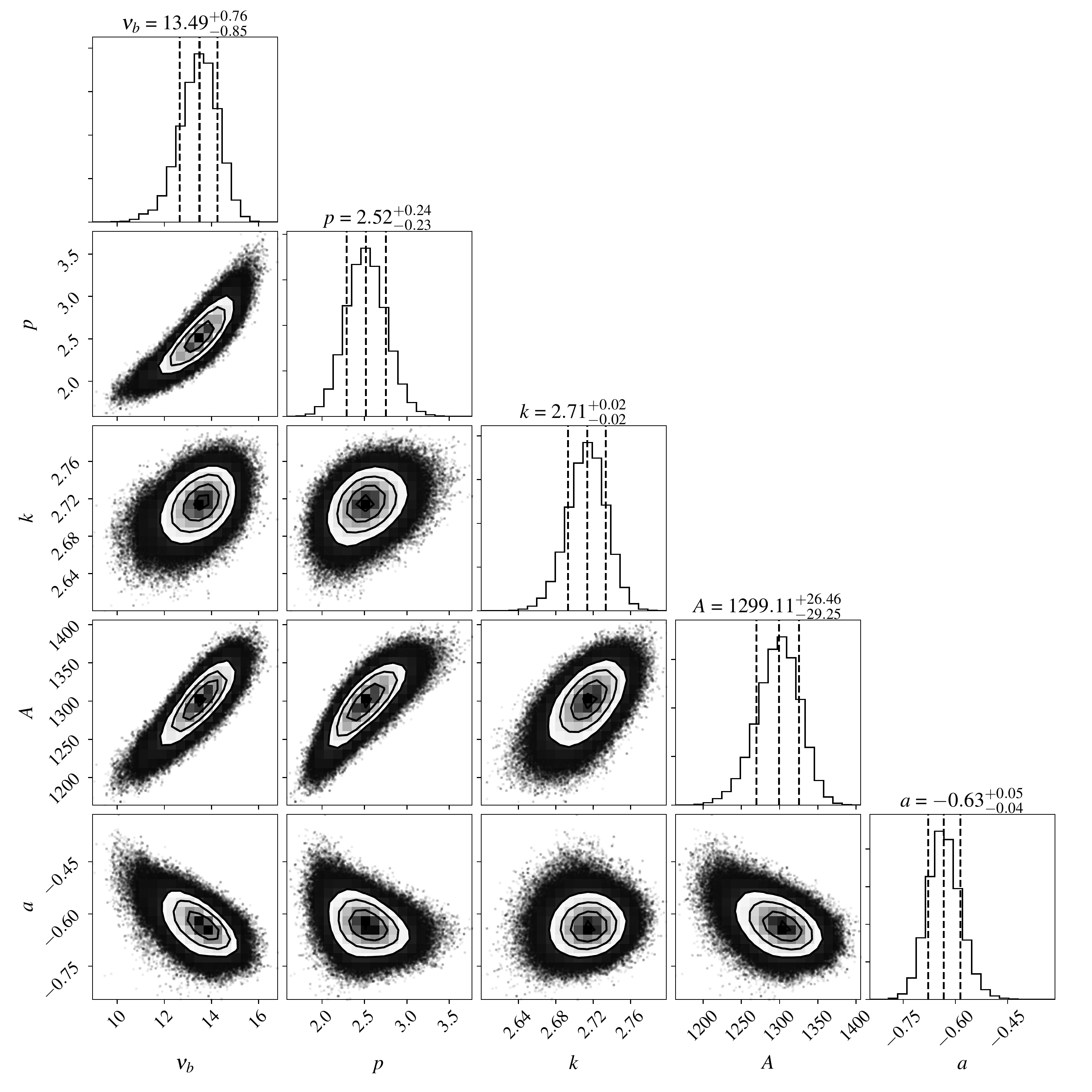}
 \caption{Marginalised posterior distribution for the joint sharply broken power law fit of the VLA epoch 2 and 3 spectra. $\nu_b$, $p$, $k$, $A$ and $a$ are the epoch 2 break frequency, electron energy index, density profile index, maximum flux and the power law index of the break frequency movement, respectively. The dotted lines indicate the $16^{\rm th}$, $50^{\rm th}$ and $84^{\rm th}$ percentiles on the parameters which are used to estimate the error.}
 \label{fig:vlaa3}
\end{figure}

We also demonstrate the effect of fitting both spectra simultaneously, this time with a sharply broken power law. To reduce the number of parameters and avoid degeneracy we fix the pre break slope in both epochs to $1/3$ as predicted by \citet{sari1998}. This leaves us with 5 variable parameters to optimise. The post-break slope (determined by $p$ as $\beta=-(p-1)/2$), the break frequency for the second epoch, the index of the power law evolution of the break, the maximum flux (the flux at the break) for the second epoch and finally the evolution of the maximum flux which depends on the density profile index as $F_{\textrm{max}}\propto t^{-k/2(4-k)}$. We opt to make the break frequency evolution a free parameter as we (unsurprisingly, given the discussion in the main text) could not find a converging fit when prescribing this to move as $-3/2$ (as it should if it is the minimum energy break) or to depend on $k$ as $-(4-3k)/2(4-k)$ if it were the cooling break. The marginalised posterior distribution for this fit is shown in Figure \ref{fig:vlaa3}.

\bsp 
\label{lastpage}
\end{document}